\documentclass[a4paper,11pt]{article}
\pdfoutput=1 
\usepackage{jcappub} 
\usepackage[T1]{fontenc} 
\usepackage{newtxtext,newtxmath}
\usepackage{ae,aecompl}
\usepackage{hyperref}
\usepackage{cleveref}

\def\prd{Physical Review D}
\def\jcap{JCAP}

\def\apj{ApJ}
\def\apjl{ApJL}

\def\aap{A\&A}

\def\mnras{MNRAS}
\def\nat{Nature}

\def\beq#1{\begin{equation}\label{#1}}
\def\eeq{\end{equation}}
\def\beqa#1{\begin{eqnarray}\label{#1}}
\def\eeqa{\end{eqnarray}}

\def\Eq#1{Eq.~(\ref{#1})} 

\def\myfrac#1#2{\left(\frac{#1}{#2}\right)}

\def\comment#1{\relax}
\def\dfrac#1#2{\displaystyle\frac{#1}{#2}}

\title{\boldmath Spins of primordial binary black holes before coalescence}

\author[a,b,c,1]{K.A. Postnov,\note{Corresponding author.}}
\author[b]{N.A. Mitichkin}

\affiliation[a]{Sternberg Astronomical Institute, M.V. Lomonosov Moscow State University,\\ 13, Universitetskij pr., 119234, Moscow, Russia}
\affiliation[b]{Faculty of Physics, M.V. Lomonosov Moscow State University,\\ Leninskie Gory, 1, 119991, Moscow, Russia}
\affiliation[c]{Department of Physics, Novosibirsk State University, \\Pirogova 2, 630090, Novosibirsk, Russia}

\emailAdd{pk@sai.msu.ru}
\emailAdd{mitichkin.nikita99@mail.ru}

\abstract{Primordial stellar-mass black holes, which may contribute to dark matter and to the observed LIGO binary black hole coalescences, are expected to be born with very low spins. Here we show that accretion mass gain by the components of a primordial black hole binary from the surrounding matter could lead to noticeable spins of the components prior to the coalescence provided high initial orbital eccentricities. 
}

\begin{document}
\maketitle
\flushbottom

\section{Introduction}
\label{s:intro}

The discovery of coalescing binary black holes (BHs) heralded the advent of gravitational wave (GW) astronomy \cite{2016PhRvL.116f1102A}. Presently, the GWTC-1 catalog of binary coalescences detected by LIGO/Virgo GW interferometers includes 10 BH+BH binaries and one NS+NS (GW170817) binary \cite{2018arXiv181112907T}. A statistical analysis of properties of coalescing binary BHs \cite{2018arXiv181112940T} suggests that the spin distribution of BHs prior to coalescence favors low spins of the components\footnote{ An independent analysis of LIGO O1 data discovered one more possible BH+BH binary, GW151216, which may have rapidly spinning aligned components, but with a low astrophysical probability $\sim 0.71$ \cite{2019arXiv190210331Z}.}. 

The origin of the observed BH binaries is not fully clear. While the evolution of massive binary systems \cite{2016Natur.534..512B,2016A&A...588A..50M,2018MNRAS.474.2959G} is able to reproduce the observed masses and effective spins of the LIGO BH+BH sources  \cite{2017arXiv170607053B,2017ApJ...842..111H,2018A&A...616A..28Q,2019MNRAS.483.3288P}, the alternative (or additional) mechanisms of the binary BH formation is not yet excluded. These channels include, in particular, the dynamical formation of close binary BH in dense stellar clusters \cite{2016PhRvD..93h4029R,2016ApJ...824L...8R} or coalescences of primordial black hole binaries which can constitute substantial fraction of dark matter \cite{1997ApJ...487L.139N,1998PhRvD..58f3003I,2016PhRvL.116t1301B,2016PhRvL.117f1101S,2016JCAP...11..036B,2016PhRvD..94h3504C,2016arXiv160404932E}. Primordial BHs may form clusters (see \cite{2019EPJC...79..246B} for a review) facilitating the formation of binary BHs. 

In this note, we focus on the last possibility in order to understand whether primordial binary BHs formed in the early Universe can have noticeable spins before the coalescence. Originally, the spins of primordial BHs should be close to zero (at a percent level at most, see e.g. recent studies \cite{2019arXiv190105963M,2019arXiv190301179D}). 
However, when in a binary system, accretion of matter will inevitably bring angular momentum, and the components of a binary BH should acquire spins.

The spin of a BH with mass $M$ and angular momentum $J$  is characterized by the dimensionless parameter $a=J/(GM^2/c)$, where $G$ and $c$ are the Newtonian gravity constant and speed of light, respectively. Below we will use geometrical units $G=c=1$. We will measure masses in solar mass units, $M_\odot=2\times 10^{33}$~g, $m=M/M_\odot$, so that the length unit is $1 [\mathrm{cm}]=2/3\times 10^{-5} m$, the time unit is $1 [\mathrm{s}]=2\times 10^5m$, etc. 

It is easy to estimate the final spin of an initially Schwarzschild BH. Assuming that no accreted mass $\Delta M=M_f-M_0$  is radiated away, the BH spin after acquiring mass $\Delta M$ reads \cite{1970Natur.226...64B}:
\beq{e:spin}
a^*=\sqrt{\frac{2}{3}}\myfrac{M_0}{M_f}\left[4-\sqrt{18\myfrac{M_0}{M_f}^2-2}\right]\,.
\eeq
This formula is valid insofar as $M_f/M_0<\sqrt{6}$. For a larger final BH mass, $a^*=a^*_{max}=1$ (more precisely, $a^*\simeq 0.998$, if one takes into account photon drag from accretion-generated radiation, \cite{1974ApJ...191..507T}).
If $\Delta M\ll M_0$, the acquired BH spin is $a*\simeq 9/\sqrt{6} (\Delta M/M_0)$.

It is easy to estimate the accretion mass gain for a single BH. Suppose it is immersed in a medium with sound velocity $c_s$. Typically, in the interstellar medium $c_s\sim \sqrt{T}\simeq 10^{-5}\sqrt{T/1 \mathrm{eV}}$ or less (here $T$ is the temperature of the medium). Assuming a Bondi-Hoyle-Lyttleton accretion onto the BH, we find 
\beq{e:dMM}
\Delta M/M_0 \approx 4 \pi \rho m/(v^2+c_s^2)^{3/2}\times t_0, 
\eeq
where $\rho$ is the density of the medium, $v$ is the proper velocity of the BH relative to the medium, $t_0$ is the duration of the accretion. For example, for the typical ISM density $\rho\sim 10^{-24}$ g~cm$^{-3}$ $=(27/16)\times 10^{-42}m^{-2}$ and a maximum possible Hubble time $t_0=t_H=4\times 10^{17} \mathrm{s}=8\times 10^{22} m$, by neglecting the BH motion, from \Eq{e:dMM} we obtain $\Delta M/M_0\simeq 1.7 10^{-3}m\ll 1$ and the final spin $a^*\simeq 3.76 \Delta M/M \simeq 0.006 m$. For a $30-50 M_\odot$ BH this would give a noticeable value but it is hard to measure the mass and spin of a single BH.

The situation is less certain for the initially non-rotating components of a binary BH that is able to coalesce over the Hubble time. Below we calculate the accretion mass gain by the components of such a binary and show that the acquired spins can be interesting only if the initial orbital eccentricity of the binary is sufficiently large. 

\section{Accretion mass gain by binary BH components}
\label{s:mass}

Consider a binary system consisting of two point-like masses $m_1$, $m_2=m_1/q$ ($q$ is the binary mass ratio). The total mass is $M=m_1+m_2=m_2(1+q)$, the orbital period $T$ is found from the 3d Kepler's law $4\pi^2/T^2=M/a^3$, where $a$ is the orbital semi-major axis.

\subsection{Circular orbits}
Let us start with the simplest case of a circular orbit. For typical BH+BH binares with $m\sim 10-50$, orbital velocities even at the maximum initial separations allowing for the coalescence over the Hubble time are much larger than the ISM sound velocity, so we will neglect $c_s$ in the Bondi-Hoyle-Lyttleton formula. 
For the $i$-th component ($i=1,2, \, j=3-i$) moving with the velocity $\upsilon_i$, the mass accretion rate (see Section \ref{s:discussion} for the discussion of the numerical coefficient) is 
\beq{e:dotMi}
\dot{M_i}=4\pi\rho \dfrac{m^2}{\upsilon_i^3}=\dfrac{4\pi\rho m_i^2a^{3/2}M^{3/2}}{m_j^3},
\eeq
where we have used the expression for the Keplerian orbital velocity of the $i$-th component $\upsilon_i=\sqrt{m_j^2/aM}$. The  binary loses the energy and angular momentum due to emission of gravitational waves, and during the time before the coalescence the mass gain by the $1$-st component (for definiteness) will to good accuracy read 
\beq{e:dM1}
\Delta M_1 = \displaystyle\int\limits_0^{t_0} \dot{M_1}dt=\displaystyle\int\limits_{a_0}^0 \frac{dM_1}{dt}\frac{dt}{da}da,
\eeq
where the initial orbital separation $a_0$ of the binary is uniquely determined from the GW-driven coalescence time
\beq{e:t0}
t_0=\dfrac{5a_0^4}{256Mm_1m_2}
\eeq
and $dt/da$ is found from the quadrupole GW formula for a circular binary system: 
\beq{e:dtda}
dt=-\dfrac{5a^3}{64m_1m_2M}da\,.
\eeq
After taking the integral in \Eq{e:dM1} and substituting $a_0$ through $t_0$ from \Eq{e:t0}, we arrive at:
\beq{e:dM1M1}
\left.\frac{\Delta M_1}{M_1}\right|_0=\dfrac{5\pi\rho M^{1/2}a_0^{11/2}}{88m_2^4}=\dfrac{5}{88}\bigg(\dfrac{256}{5}\bigg)^{11/8}\pi\rho t_H^{11/8}m_1^{5/8}q^{3/4}(1+q)^{15/8}.
\end{equation}

\begin{figure}
	\includegraphics[width=\columnwidth]{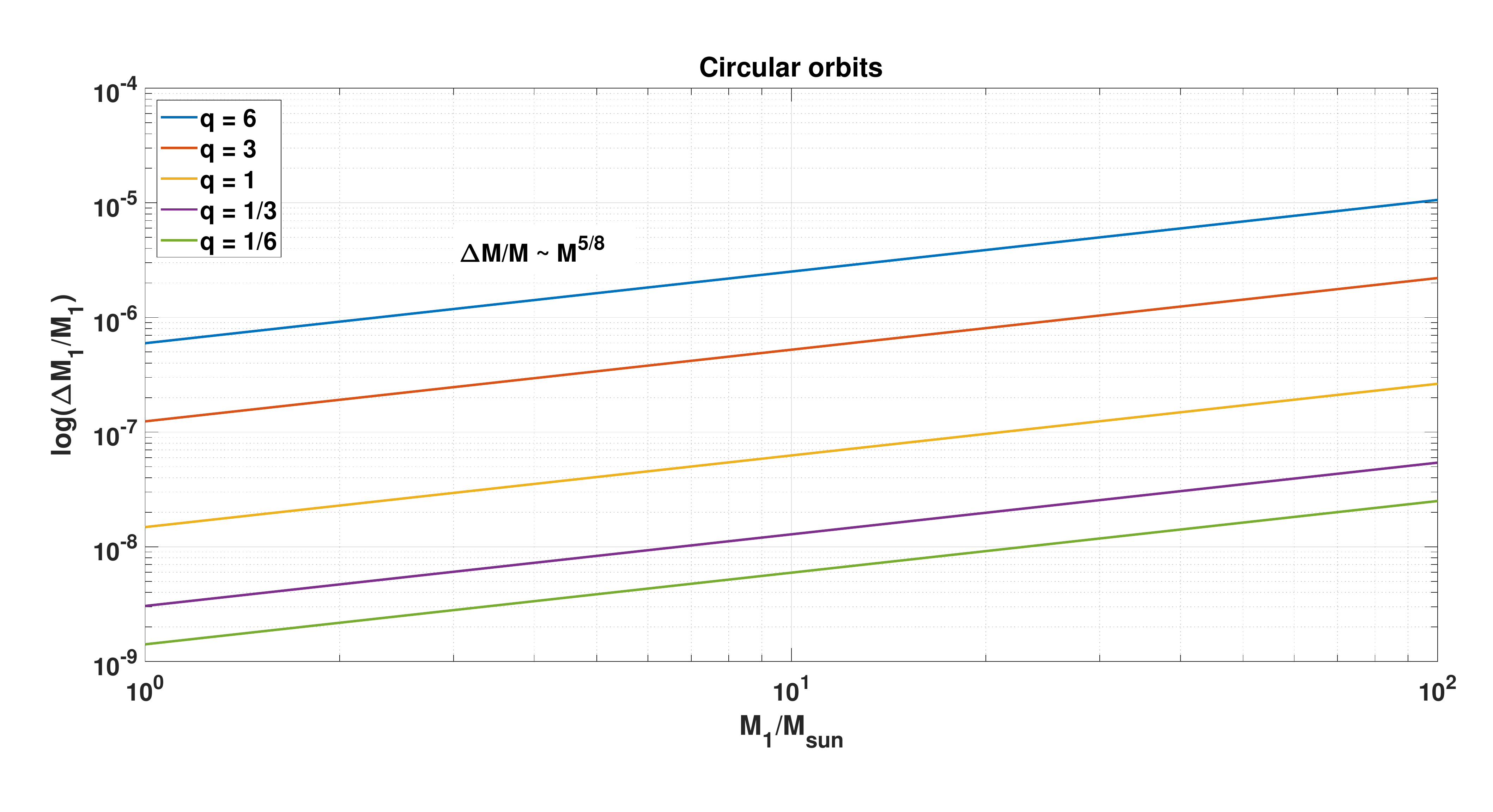}
    \caption{The fractional accretion mass gain by a coalescing binary BH system in a circular orbit over the Hubble time in a cold medium with density  1 cm$^{-3}$.}
    \label{f:circular}
\end{figure}

The plot of $\Delta M_1/M_1$ as a function of $m_1$ is shown in Fig. \ref{f:circular} for different mass ratios $q=m_1/m_2$ for the fiducial ISM density 1 g~cm$^{-3}$. Clearly, the effect increases both with $m_1$ and $q$ but even for large $q>1$ (i.e., when we consider the mass gain by the heaviest binary component) is desperately small to enable astrophysically interesting BH spins, even for larger densities.

\subsection{Elliptical orbits}

The case of initially elliptical orbits is more interesting. Elliptical orbits of binary BHs are possible in both the dynamical channel of binary BH formation in dense stellar clusters and for primordial BHs. 

Consider a Keplerian binary in an elliptical orbit with eccentricity $e_0$. The mass accreted over one orbital revolution with period $T$ reads:
\beq{e:deltaM}
\delta M_1 = \int\limits_0^T\dot M_1 dt=2\int\limits_0^\pi \dot M_1\myfrac{dt}{d\theta} d\theta = 8\pi\rho q^2(1+q)(1-e^2)a^3  I_1(e)
\eeq
where 
\beq{e:I1}
I_1(e)=\displaystyle\int\limits_0^{\pi}\bigg((1+e\cos{\theta})^2(1+2e\cos{\theta}+e^2)^{3/2}\bigg)^{-1}d\theta\,. 
\eeq
Here we have used the expressions for the orbital velocity $\upsilon_i(\theta)=\sqrt{M/a(1-e^2)}(1+2e\cos\theta+ e^2)^{1/2}(m_j/M)$, the orbital angular momentum conservation  $r^2d\theta/dt=\sqrt{Ma(1-e^2)}$ and $r=a(1-e^2)/(1+e\cos\theta)$ for the Keplerian motion. 

The mass accretion rate averaged over one orbital period $T$ is 
\beq{e:avdM1}
\langle\dot{M_1}\rangle = \frac{\delta M_1}{T}, \quad
T = 2\pi\sqrt{\dfrac{a^3}{M_2(1+q)}}\,. 
\eeq
In a way similar to the circular case, we find the accretion mass gain by the 1-st component of a binary BH with initial orbital eccentricity $e_0$ coalescing over the Hubble time:
\beq{e:dMe}
\Delta M_1(e_0) = \displaystyle\int\limits_{e_0}^0\langle\dot{M_1}\rangle\bigg(\dfrac{dt}{de}\bigg) de\,,
\eeq
where $de/dt$ reads \cite{1964PhRv..136.1224P}
\beq{e:dedt}
\dfrac{de}{dt} = -\dfrac{304m_1m_2M  e}{15a^4(1-e^2)^{5/2}}\bigg(1+\dfrac{121}{304}e^2\bigg)\,.
\eeq

For the coalescing binary, the expression $a(e)$ reads \cite{1964PhRv..136.1224P} 
\beq{e:ae}
a(e)=\dfrac{C_0e^{12/19}}{(1-e^2)}\bigg(1+\dfrac{121}{304}e^2\bigg)^{870/2299}\,,
\eeq
where the constant $C_0$ is determined by substituting $a(e_0)$ into the formula for the binary coalescence time $t_0$ in the case of elliptical orbit \cite{1964PhRv..136.1224P}:
\beq{e:t0e}
t_0=\dfrac{5a(e_0)^4}{256Mm_1m_2} \dfrac{48(1-e_0^2)^4}{19e_0^{48/19}}\bigg(1+\dfrac{121}{304}e_0^2\bigg)^{-3480/2299}  I_2(e_0)\,, \quad
I_2(e_0) = \displaystyle\int\limits_0^{e_0}\dfrac{\bigg(1+\dfrac{121}{304}e^2\bigg)^{1181/2299}  e^{29/19}}{(1-e^2)^{3/2}}de\,. 
\eeq

\begin{figure}
	\includegraphics[width=0.49\textwidth]{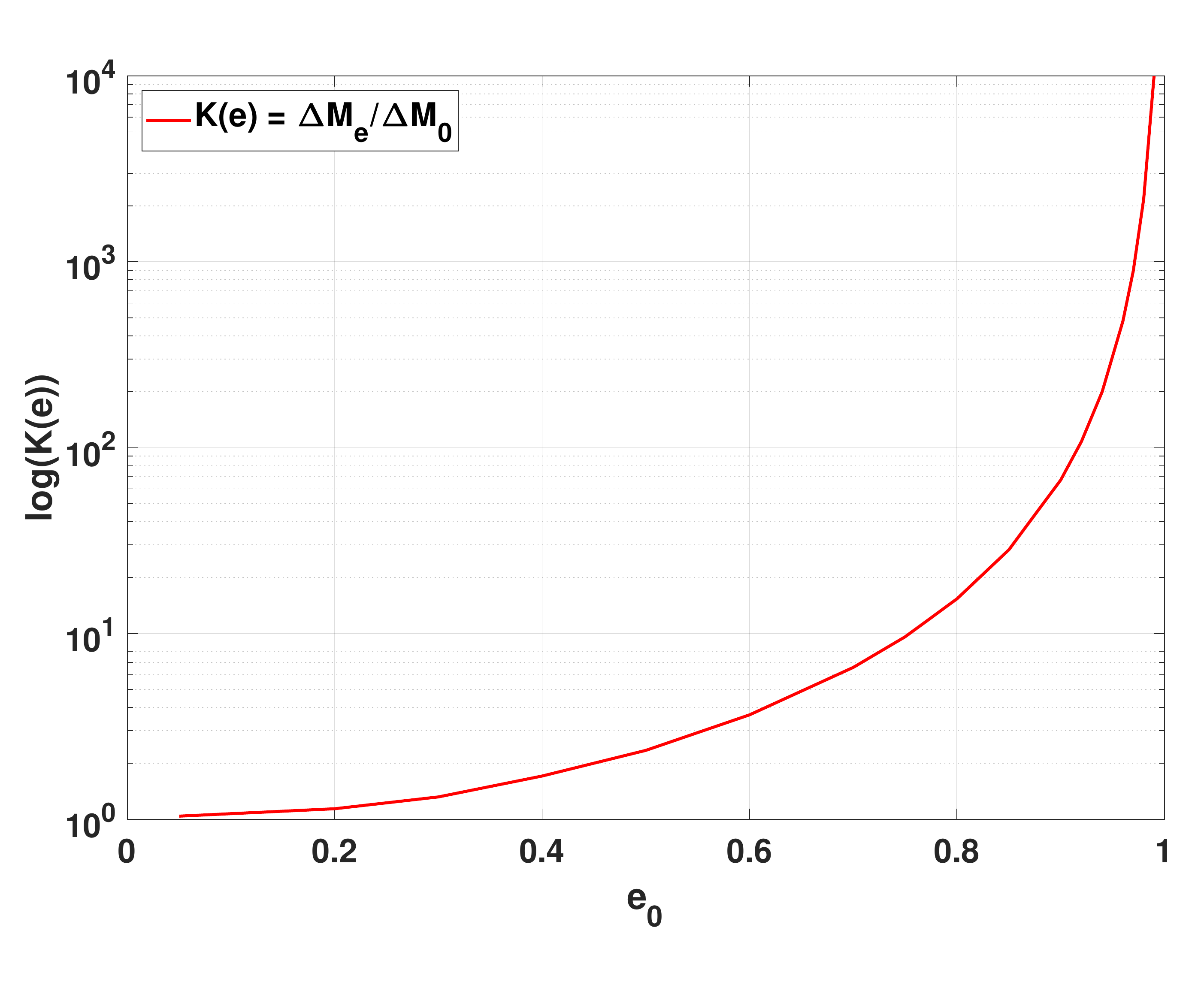}
	\includegraphics[width=0.49\textwidth]{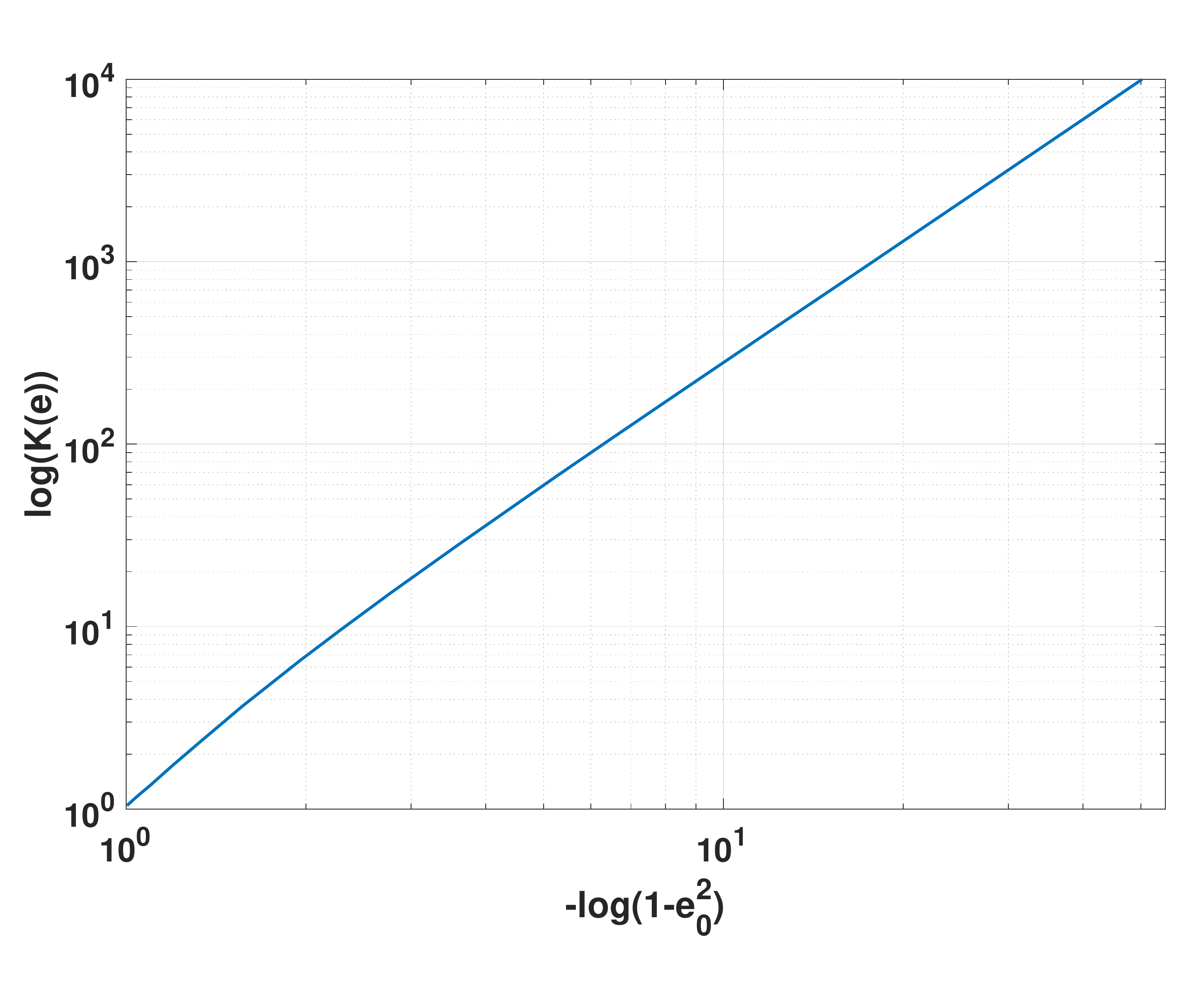}
    \caption{Left: The enhancement factor $K(e_0)$ of the fractional mass accretion gain in elliptical orbit by a component of a coalescing binary BH relative to the circular case as a function of the orbit eccentricity $e_0$. Right: $\log K(e_0)$ - $(-\log(1-e_0)^2)$ plot manifestly showing the asymtptotic power-law behaviour at large $e_0\simeq 1$, $K(e_0)\sim (1-e_0^2)^{-2.58}$.}
    \label{f:Ke}
\end{figure}

Substituting \Eq{e:avdM1} into \Eq{e:dMe} with an account of \Eq{e:ae} and \Eq{e:t0e}, we finally obtain the accretion mass gain for the elliptical orbit:
\beq{e:dMeM1}
\left.\frac{\Delta M_1}{M_1}\right|_e=\dfrac{15}{76}\bigg(\dfrac{304}{15I_2(e_0)}\bigg)^{11/8}\rho t_H^{11/8}m_1^{5/8}q^{3/4}(1+q)^{15/8}
\displaystyle\int\limits_0^{e_0}
I_1(e)e^{47/19}\bigg(1+\dfrac{121}{304}e^2\bigg)^{226/209}de.
\eeq
It can be written in the form 
\begin{equation}
\left.\frac{\Delta M_1}{M_1}\right|_e=\left.\frac{\Delta M_1}{M_1}\right|_0\cdot K(e_0),
\end{equation}
where the enhancement factor $K(e_0)$ reads:
\beq{e:Ke}
K(e_0)=\frac{66}{19\pi} \bigg(\frac{19}{48}\frac{1}{I_2(e_0)}\bigg)^{11/8}
\displaystyle\int\limits_0^{e_0}
I_1(e)e^{47/19}\bigg(1+\dfrac{121}{304}e^2\bigg)^{226/209}de.
\eeq
Clearly, the enhancement factor with respect to the circular orbit is a function of the initial orbital eccentricity only, and is shown in Fig. \ref{f:Ke}, left panel. In the $e_0\to 0$ limit, $I_1(e_0)\sim \pi$ and $I_2(e_0)\sim (19/48)e_0^{48/19}$, and $K_{e_0}\to 1$. In the more interesting limit of large eccentricities $e_0\simeq 1$, we find from numerical integration $I_1(e_0)\sim (1-e_0^2)^{-4.27}$. Therefore, in this limit $K(e_0)\sim (1-e_0^2)^{11/16}\times (1-e_0^2)^{-3.27}=(1-e_0^2)^{-2.58}$. This power-law asymptotic is clearly seen on the plot $\log K(e_0)-\log 1/(1-e_0^2)$ shown in Fig. \ref{f:Ke}, right panel. Therefore, in the limit of high initial orbital eccentricities, we find approximately  
\begin{eqnarray}
\label{e:dMM1}
\left.\frac{\Delta M_1}{M_1}\right|_e &\approx 10^{-5}
\myfrac{\rho}{10^{-24} \mathrm{g\,cm}^{-3}}\myfrac{M_1}{30 M_\odot}^{5/8}
q^{3/4}(1+q)^{15/8}\myfrac{0.1}{1-e_0^2}^{2.58}\nonumber \\
&\approx 10^{-5}
\myfrac{\rho}{10^{-24} \mathrm{g\,cm}^{-3}}\myfrac{\cal M}{30 M_\odot}^{5/8}
q(1+q)^2\myfrac{0.1}{1-e_0^2}^{2.58}\,.
\end{eqnarray}
In the last equality, we have introduced the chirp mass of the binary system  ${\cal M}\equiv (M_1M_2)^{3/5}/M^{1/5}=M_1(q^2(1+q))^{-1/5}$ that is directly read off the chirp GW signal from binary coalescences. Note that \Eq{e:dMM1} does not violate the isolated BH mass gain estimate, $\Delta M/M\simeq 1.7 \times 10^{-3}m$ for the orbital eccentricities $1-e_0^2>0.0037 (m_1/30)^{0.24}[q^{3/4}(1+q)^{15/8}]^{1/2.58}$, i.e. $e_0<0.997$ for $m_1=30$ and $q=1$. This limit, however, is weaker than imposed by the validity of negligible sound velocity in the Bondi-Hoyle-Lyttleton formula (see the next Section).  

\begin{figure}
	\includegraphics[width=\columnwidth]{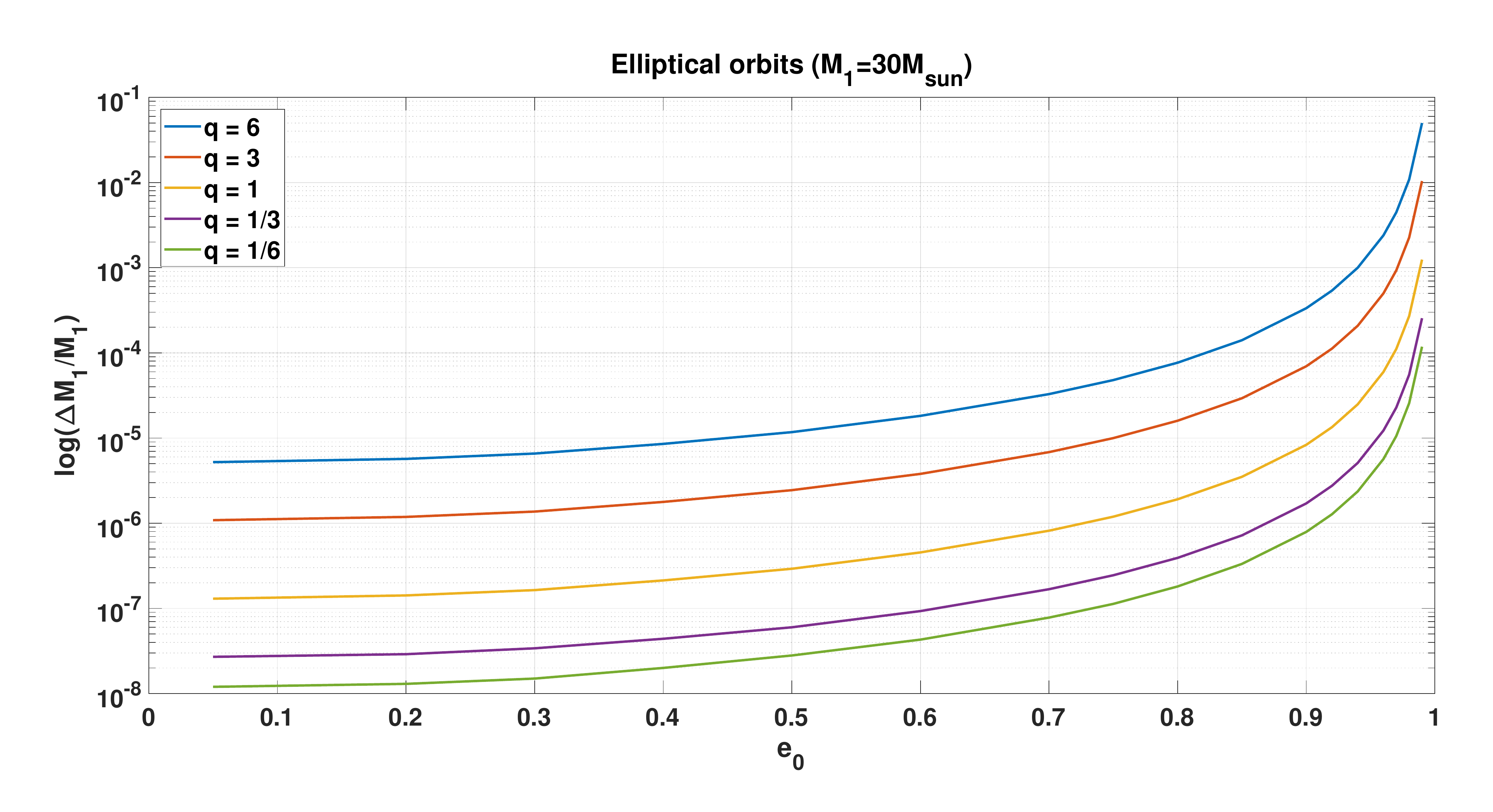}
    \caption{The fractional accretion mass gain by a $30 M_\odot$ black hole in an elliptical binary as a function of the initial orbital eccentricity $e_0$. The cold medium density is 1 cm$^{-3}$.}
    \label{f:elliptic}
\end{figure}

It is seen that for eccentric orbits with $e_0\gtrsim 0.95$ (i.e. $1/(1-e_0^2)\gtrsim 10)$) this factor can bring the mass accretion gain into astrophysically interesting region for BH spin, especially for more massive component of a binary with large mass ratio $q>1$.  As an example, in Fig. \ref{f:elliptic} we show the fractional accretion mass gain by a BH with mass $m_1=30$ as a function of the initial orbital eccentricity $e_0$ for different binary mass ratios $q$. Formally, for the assumed ISM density, a noticeable spin of the primary BH component before the coalescence, $a^*\sim 3.76 (\Delta M_1/M_1)$, could be achieved only for very eccentric orbits with $e_0\simeq 1$. The effect is stronger for more massive BHs and higher surrounding densities.

\section{Discussion} 
\label{s:discussion}

\begin{figure}
	\includegraphics[width=\columnwidth]{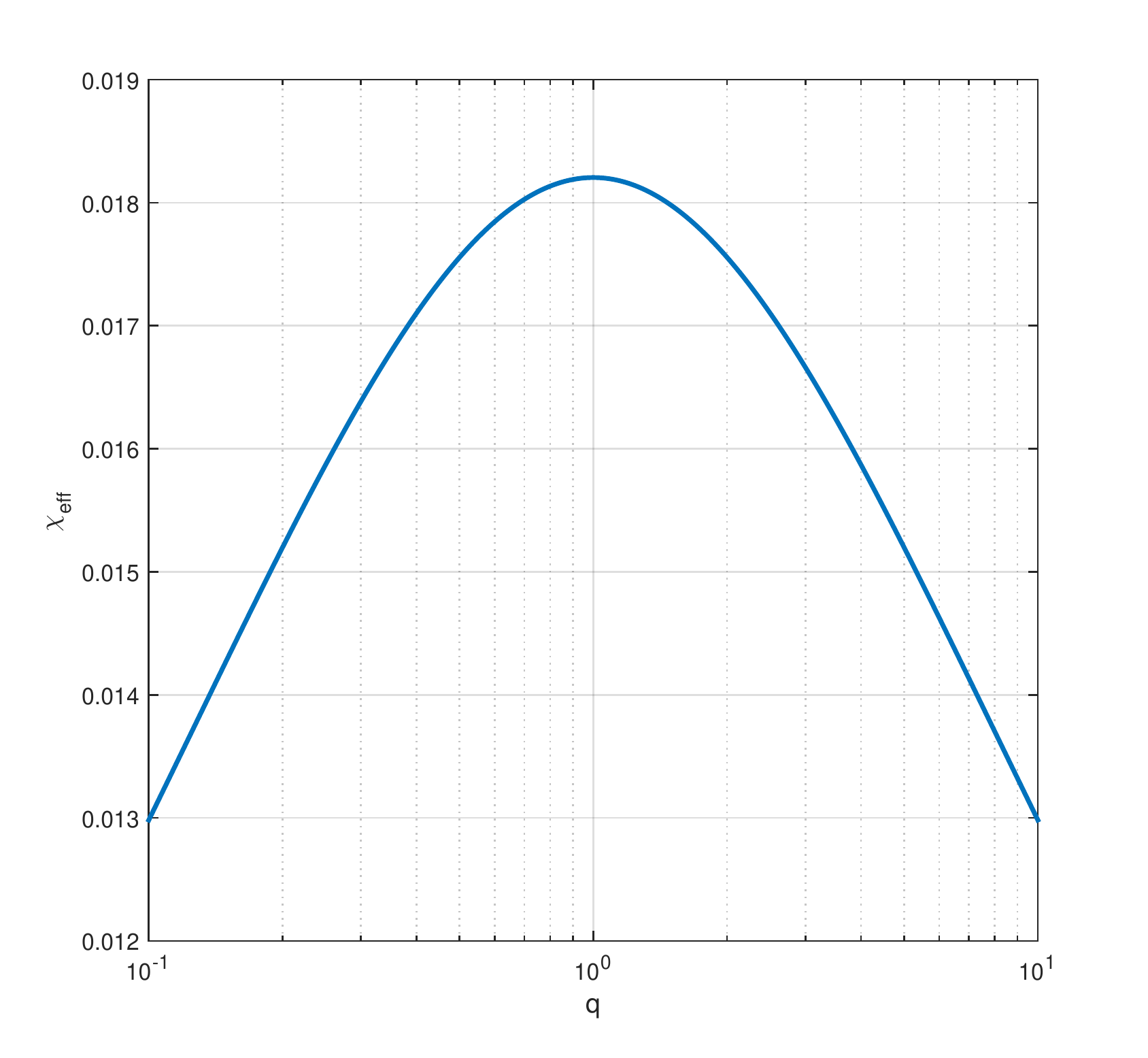}
    \caption{Maximum possible effective spin of a coalescing BH with chirp mass ${\cal M}=30 M_\odot$ that had accreted matter from cold ISM ($c_s=10^{-5}$) with particle number density 1 cm$^{-3}$.}
    \label{f:chimax}
\end{figure}

In our analysis, we have neglected the sound velocity $c_s^2$ in the Bondi-Hoyle-Lyttleton formula \Eq{e:dMM}. This needs to be justified in the case of strongly eccentric orbits because the accretion rate is determined by the maximal of the orbital velocity and the sound velocity $c_s$. The orbital velocity of star $M_1$ at  apastron is
\beq{e:v1p}
v_a(M_1)=\sqrt{\frac{M(1-e)^2}{a(1-e^2)}}\frac{M_2}{M}= \sqrt{\frac{M_1(1-e^2)}{a(1+e)^2q(1+q)}}\,.
\eeq
Then the condition $v_a(M_1)>c_s$ can be written as
\beq{e:vgtcs}
1-e^2>c_s^2(1+e)^2q(1+q)\myfrac{a}{M_1}\,.
\eeq
Clearly, if the initial orbital eccentricity $e_0$ satisfies this inequality, it will hold always true in the subsequent binary evolution due to GW losses. 
Making use of \Eq{e:t0e} to express $a(e_0)/M_1$, 
in the limit $e_0\to 1$ of interest here we find 
\[
1-e_0^2> 4c_s^2q^{5/4}(1+q)^{5/4}\bigg(\frac{256}{5}\frac{t_H}{m_1}\bigg)^{1/4}\bigg(\frac{19}{48(1-e_0^2)^4}\bigg)^{1/4}\bigg(\frac{425}{304}\bigg)^{870/2299}[I_2(e_0\to 1)]^{-1/4}
\]
Noticing that $I_2(e_0\to 1)\approx \bigg(\frac{425}{304}\bigg)^{1181/2299}(1-e_0^2)^{-1/2}$, plugging $t_H/m_1$ for the fiducial $m_1=30$, after making arrangements, we arrive at the inequality
\beq{e:e01max}
1-e_{0,\mathrm{max}}^2(m_1)>0.01\myfrac{c_s}{10^{-5}}^{16/15}\myfrac{m_1}{30}^{-2/15}(q(1+q))^{2/3}
\eeq
that restricts the applicability of our approximation. 
Proceeding exactly in the same way as for $m_1$, we obtain the restriction for the initial orbital eccentricity for $m_2$:
\beq{e:e02max}
1-e_{0,\mathrm{max}}^2(m_2)>0.01\myfrac{c_s}{10^{-5}}^{16/15}\myfrac{m_1}{30}^{-2/15}q^{-6/5}(1+q)^{2/3}\,.
\eeq

This limit for the initial binary eccentricity, $e_{0,\mathrm{max}}<1-0.005$ for the fiducial parameters, leaves quite a room for a significant enhancement factor $K(e_0)$. To see this, let us estimate the maximum possible effective spin of a coalescing binary BH, $\chi_{\mathrm{eff}}=(m_1a^*_1+m_2a^*_2)/M$, which can be inferred from GW observations \cite{2018arXiv181112907T}. Note that in our setup fully aligned BH spins are expected. As $a^*_i\sim (\Delta M/M)_i$, we need to calculate also the mass gain by the secondary component, $(\Delta M/M)_2$. This is obviously done by substituting $M_1
\to M_2$ in \Eq{e:dMM1}, i.e. simply changing $q\to 1/q$: 
\beq{e:dMM2}
\frac{\Delta M_2}{M_2}\approx 10^{-5}
\myfrac{\rho}{10^{-24} \mathrm{g\,cm}^{-3}}\myfrac{\cal M}{30 M_\odot}^{5/8}
q^{-3}(1+q)^2\myfrac{0.1}{1-e_0^2}^{2.58}\,.
\eeq

If there would be no initial eccentricity restrictions (e.g., in the limit of a cold medium with very low sound velocities $c_s$), the effective spin of the coalescing BH binary with ${\cal M}=30 M_\odot$ would be
\begin{eqnarray}
\label{e:chieff}
\chi_{\mathrm{eff}}&=\frac{q}{1+q}a^*_1+\frac{1}{1+q}a^*_2\approx 
3.76\times 10^{-5}\myfrac{\rho}{10^{-24} \mathrm{g\,cm}^{-3}}\myfrac{\cal M}{30 M_\odot}^{5/8}
\myfrac{0.1}{1-e_0^2}^{2.58}(1+q)(q^2+q^{-3})\nonumber\\
&>5.3\times 10^{-4}\myfrac{\rho}{10^{-24} \mathrm{g\,cm}^{-3}}\myfrac{\cal M}{30 M_\odot}^{5/8}
\myfrac{0.1}{1-e_0^2}^{2.58}
\end{eqnarray}
for any mass ratio $q$ because the function $f(q)=(1+q)(q^2+q^{-3})$ reaches the minimum  $f(q_{min})=4$ at $q_{min}=1$. However, taking into account the initial eccentricity limits, \Eq{e:e01max} and \Eq{e:e02max}, due to finite sound velocity of the medium, we find 
\beq{e:chieffmax}
\chi_{\mathrm{eff}}<\chi_{\mathrm{eff},\mathrm{max}}=\frac{q}{1+q}a^*_1(e_{0,\mathrm{max}}(m_1))+\frac{1}{1+q}a^*_2(e_{0,\mathrm{max}}(m_2))\propto {\cal M}^{0.97}c_s^{-2.75}\Psi(q)\,,
\eeq
where $\Psi(q)$ is a symmetric function of the mass ratio with maximum at $q=1$ that can be readily calculated by substituting the factors $1-e_{0,\mathrm{max}}^2(m_{1,2})$ for $a_1$ [\Eq{e:e01max}] and $a_2$ [\Eq{e:e02max}], respectively, into \Eq{e:chieff}. The plot of $\chi_{\mathrm{eff},\mathrm{max}}$ for the fiducial values $c_s=10^{-5}$, $\rho=10^{-24}$ g cm$^{-3}$ and ${\cal M}=30 M_\odot$ is shown in Fig. \ref{f:chimax}. Roughly, we can take $\chi_{\mathrm{eff},\mathrm{max}}\simeq 0.01(\rho/10^{-24}\hbox{g\,cm}^{-3})({\cal M}/30\,M_\odot)^{0.97}(c_s/10^{-5})^{-2.75}$ for any mass ratio $0.1<q<10$. This estimate shows that coalescing primordial binary BHs can acquire measurable values of $\chi_{\mathrm{eff}}\sim $ a few percents due to the accretion mass gain in the galactic ISM.

\section{Conclusion}

Here we presented the results of  calculation of the mass gain by components of a BH+BH binary system, which can coalesce over the Hubble time, due to the Bondi-Hoyle-Lyttleton accretion from a relatively cold ($c_s\sim$ a few km s$^{-1}$) surrounding medium. The angular momentum by the accreted material can spin up an the initially Schwarzschild BH up to noticeable values if the initial binary orbit had a high eccentricity $e_0\sim 1$. Such eccentricities are in principle possible in primordial BH binaries that can be formed in the early Universe and coalesce at the present time. 

In our calculations we have assumed the simplest formula for the accretion rate onto a binary components, which is, of course, a rough estimate. For example, recent 3D simulations of Bondi-Hoyle accretion \cite{2019arXiv190107572A} suggest an orbital-averaged reduction of the Bondi-Hoyle accretion efficiency by a factor of $\sim 1/4$ in circular binaries. However, for our purposes this reduction is not very important in view of much more uncertain density of matter surrounding the coalescing binary. This density can be an order of magnitude higher or smaller depending on the location of the binary in a galaxy or in the galactic halo. Moreover, primordial binary BHs are thought to have high velocity dispersion $\sim 300\,\mathrm{km\,s}^{-1}$, which drastically reduces the efficiency of matter accretion. Still, some BH binaries could have rather small velocities and can be found inside the galactic ISM. Therefore, in principle, the components  of such BH binaries can acquire noticeable aligned spins prior to the coalescence. 

The accretion-gained spins should be higher in more massive binaries ($a^*\sim \Delta M/M_0\sim m^{5/8}$, see Fig. \ref{f:circular}). Interestingly, the most massive LIGO binary BH, GW170729 \cite{2018arXiv181112907T} and (not very reliable)  recently reported BH binary GW151216 \cite{2019arXiv190210331Z} show appreciable and likely aligned spins of the components. Of course, presently it is difficult to separate different formation channels of the observed coalescing binary BHs, and increased statistics of binary BH coalsecences in the forthcoming O3 LIGO/Virgo run could help disentangling various scenarios of binary BH formation and evolution. However, we stress that even primordial binary BHs could have appreciable aligned spins before the coalescence due to matter accretion in galaxies. 

\acknowledgments

We thank the anonymous referee for constructive criticism and Prof. A.D. Dolgov for encouraging discussions.
The work of KAP is supported by RSF grant 19-42-02004. 
NAM acknowledges support by the Program of development of M.V. Lomonosov Moscow State University (Leading Scientific School 'Physics of stars, relativistic objects and galaxies').




\bibliographystyle{JHEP}
\bibliography{PBH} 
\end{document}